\documentstyle[amssymb,aps,prl,twocolumn,epsf]{revtex}

\begin{document}
\draft

\flushbottom
\twocolumn[\hsize\textwidth\columnwidth\hsize\csname
@twocolumnfalse\endcsname
\title{CDW, Superconductivity and Anomalous Metallic Behavior in 2D Transition Metal Dichalcogenides}
\author{A.~H.~Castro Neto$^*$}
\address
{Department of Physics, Boston University, Boston, MA 02215 }
\date{\today }
\maketitle
\widetext\leftskip=1.9cm\rightskip=1.9cm\nointerlineskip\small
\begin{abstract}
\hspace*{2mm}
We propose a theory for quasi-two-dimensional transition metal dichalcogenides that provides a unified microscopic picture of the charge density wave (CDW) and superconducting phases. We show, based on the electron-phonon coupling and Fermi surface topology, that a CDW order parameter with six-fold symmetry and nodes (f-wave) gives a consistent description of the available experimental data. The elementary excitations in the CDW phase are Dirac electrons. The superconducting state has its origin on the attractive interaction mediated by phonons. The theory predicts strong deviations from Fermi liquid theory in the CDW phase. 
\end{abstract}
] \narrowtext

The quasi-two-dimensional (2D) transition metal dichalcogenides (TMD), 2H-TaSe$_2$, 2H-TaS$_2$, 2H-NbSe$_2$ and 2H-NbS$_2$ have been extensively studied since their discovery \cite{review}. These are layered systems with a phase diagram where CDW order co-exists with superconductivity at low temperatures (see Fig.\ref{phasediagram}). Many physical properties such as  resistivity, thermal expansion, sound velocity and so on, are very anisotropic and differ by orders of magnitude if measured parallel or perpendicular to the metallic planes \cite{exp1,exp2}. Neutron scattering studies have shown that these systems undergo a triple incommensurate CDW transition at a temperature $T_{cdw}$
with wave-vectors ${\bf Q}_i = (1-\delta) {\bf b}_i/3$ where
${\bf b}_i$ are the three reciprocal lattice vectors for a triangular lattice with lattice spacing $a$ ($|{\bf b}_i|=4 \pi/(\sqrt{3} a)$) and $\delta \ll 1$ is the incommensurability (TaSe$_2$ also has a 
transition into a commensurate state, $\delta =0$) \cite{neutrons}.

Although the phenomenological description of the CDW transition is well understood due to the work of McMillan \cite{mac} the current understanding of the microscopic origin of the CDW state is not complete. Fermi surface nesting, which is the main mechanism for CDW formation in 1D systems (such as 1T-TaSe$_2$ \cite{exp2}) leads to a lattice deformation with momentum $2 k_F$ (where $k_F$ is the Fermi momentum), to the gapping of the Fermi surface and therefore to insulating behavior. Band structure calculations (BSC) \cite{review,wexler} do not show, however, strong nesting at the Fermi surface of TMD. Moreover, TMD become better metals (with a resistivity drop \cite{exp2}) inside of the CDW phase and the specific heat behaves like a power of the temperature, $T$, (instead of exponentially as in the case of an 1D CDW transition). What is more striking is that TMD become superconductors at low temperatures. An alternative scenario was proposed by Rice and Scott (RS) \cite{rice} based on earlier BSC \cite{mat} in which electron scattering with momentum ${\bf Q}_{SP}$ between different saddle-points (which produce a logarithmically divergent density of states) would drive the system into the CDW phase. The advantages of this picture are clear: it does not require nesting, leads to a gap at the saddle points and to the reduction of the resistivity. More refined BSC have shown that the saddle points are not as close to the Fermi surface as initially believed \cite{wexler} (see Fig.\ref{fermisurface}).

With the advent of high temperature superconductors (HTC) many experimental techniques have been perfected. Among them, angle resolved photoemission (ARPES) has nowadays momentum and energy resolution that was unavailable when TMD were discovered. ARPES measures the electronic spectral function, the loci of the Fermi surface, and the momentum, ${\bf k}$, and frequency, $\omega$, dependent electron self-energy, $\Sigma({\bf k},\omega)$. The imaginary part of $\Sigma({\bf k},\omega)$ provides a direct measurement of the electron lifetime and for a Fermi liquid behaves like $\Im[\Sigma(k_F,\omega \to 0)] \propto 1/\tau_0 + b \, \omega^2/E_F$, where $\tau_0$ is the impurity scattering lifetime, $E_F$ is the Fermi energy and $b$ is a constant (we use units such that $\hbar=k_B=1$). Recent measurements of $\Sigma ({\bf k},\omega)$ for HTC \cite{htc} have shown that $\Im[\Sigma (k_F,\omega)] \propto |\omega|$ in clear disagreement with Fermi liquid theory but in agreement with the marginal Fermi liquid description \cite{mfl}. Stimulated by the recent developments in ARPES, TMD have been extensively studied in the last two years. As a result of these studies a contradictory picture of the origin of the CDW state has emerged. Although there are indications for the opening of a CDW gap along the saddle point direction (but in no other place along the Fermi surface) \cite{liu}, the measured $|{\bf Q}_{SP}|$ is larger than $2 k_F$ and $|{\bf Q}_i|$ (see Fig.\ref{fermisurface}) \cite{straub}. Furthermore, ARPES measurements of the electronic self-energy have shown that, analogously to the case of HTC, $\Im[\Sigma(k_F,\omega \to 0)] \propto  1/\tau + b \, |\omega|$ \cite{kevin}. These measurements clearly show that the current theories for CDW formation cannot describe the current experimental situation.

We propose that, due to strong variations in the electron-phonon coupling and to imperfect nesting of the Fermi surface, the CDW gap is six-fold symmetric and has nodes (f-wave). This f-wave state has lobes (the largest value of the CDW gap, $Max[|\Delta_{cdw}({\bf k})|]$) along the saddle point direction. This effect leads to the reduction of the electron-phonon scattering in the same way as the RS mechanism. The low-lying excitations in the CDW state are Dirac electrons associate with the nesting between different branches (around the $\Gamma$ and $K$ points) of the Fermi surface which are connected by ${\bf Q}_i$. Thus, the system remains a good metal in the CDW phase. Due to the loss of lattice inversion symmetry in the CDW phase, the Dirac electrons are coupled via a piezo-electric coupling to acoustic phonons. This coupling produces a damping of the Dirac electrons and leads to a $\Sigma({\bf k},\omega)$ of the form ($T \ll T_{cdw}$): 
\begin{eqnarray}
\Im[\Sigma(k_F,\omega \to 0)] = \tau_0^{-1}+ \alpha \, T +  \gamma \, |\omega|
\label{marginal}
\end{eqnarray}
for $\omega < Max[|\Delta_{cdw}|]$ and $\Im[\Sigma(k_F,\omega \to 0)] = 1/\tau_1$, constant, for $Max[|\Delta_{cdw}|] < \omega$. We show that the data is in good agreement with these results. Finally, we argue that phonons drive the system to a superconducting state via a Kosterlitz-Thouless (KT) phase transition. We propose that critical fluctuations of the superconducting 
order parameter, $\Delta_s$, lead to a pseudo-gap behavior above the transition temperature,
$T_c$, to a drop of the magnetic susceptibility, and to diamagnetic response above $T_c$ (as seen experimentally \cite{geballe}). 

Unlike ordinary metals, the electron-phonon coupling, $g_{\lambda}({\bf k},{\bf q})$, (where 
${\bf q}={\bf k'}-{\bf k}$ is the momentum transfer for an electron being scattered by a phonon from ${\bf k}$ to ${\bf k'}$ and $\lambda$ is the phonon polarization) in transition metals is highly dependent on the electron momentum. This dependence is responsible for many anomalies observed in the phonon spectrum in these systems \cite{varma}. Numerical studies have shown that $g_{\lambda}({\bf k},{\bf q})$ can vanish at certain points in the Brillouin zone \cite{motizuki}. A simple tight-binding calculation gives,
\begin{eqnarray}
g_{\lambda}({\bf k},{\bf q}) &=& \sqrt{\frac{2}{N M \omega_{\lambda,{\bf q}}}}
\sum_{{\bf n}} c_{|{\bf n}|} {\bf e}_{\lambda,{\bf q}} \cdot
{\bf n} \sin({\bf q}\cdot {\bf n}/2) \times
\nonumber
\\
&\times& \cos[({\bf k}+{\bf q}/2) \cdot {\bf n}] \, , 
\label{gkq}
\end{eqnarray} 
where $N$ is the number of atoms, $M$ is the atom mass, $\omega_{\lambda,{\bf q}}$ is the phonon frequency, ${\bf e}_{\lambda,{\bf q}}$ the polarization vector (${\bf e}^*_{\lambda,-{\bf q}} = {\bf e}_{\lambda,{\bf q}}$), ${\bf n}$ the nearest neighbor vector and $c_{|{\bf n}|}$ depends only on $|{\bf n}|$. It is easy to see that $g_{\lambda}({\bf k},{\bf Q}_1)$ vanishes along the nodal lines in Fig.\ref{fermisurface}. BSC (and ARPES measurements) show that the Fermi surface is hole-like, centered around the $\Gamma$ and $K$ points, and intercepts the nodal lines at the Dirac points (see Fig.\ref{fermisurface}). 
At these points the CDW order parameter vanishes since 
$\Delta_{cdw}({\bf k}) \propto g_{\lambda}({\bf k},{\bf Q}_i)$. Observe that ${\bf Q}_1$ connects nodal points located at different branches of the Fermi surface. Moreover, as shown by numerical calculations \cite{wexler} these points are well nested and a CDW state can emerge. In a triple CDW phase, however, the situation is more complicated because ${\bf Q}_2$ connects the nodal points in the same branch of the Fermi surface. This would imply a finite coupling at these points and the opening of a gap there as well. We observe, however, that in this direction the Fermi surface is not nested \cite{wexler} in contrast with the case discussed above (these details are not shown in Fig.\ref{fermisurface}). Thus, a gap cannot open at the Dirac points. The conclusion is that the CDW gap will change sign in six points along the Fermi surface leading to the situation depicted on Fig.\ref{fermisurface}. 
This is the f-wave CDW state.

Let $c_{{\bf k},\sigma}$ ($c^{\dag}_{{\bf k},\sigma}$) be electron annihilation (creation) 
operator for an electron with momentum ${\bf k}$ and spin projection $\sigma$ ($\sigma= 
\uparrow,\downarrow$). We can define the spinor operators
\begin{eqnarray}
\Psi^{\dag}_{i,\sigma}({\bf k}) = (c^{\dag}_{{\bf k},\sigma},c^{\dag}_{{\bf k}+{\bf Q}_i,\sigma})= (\psi^{\dag}_{+,i,\sigma}({\bf k}), \psi^{\dag}_{-,i,\sigma}({\bf k}))
\label{spinor}
\end{eqnarray}
where $+,-$ indicates if the fermion is particle or hole (anti-particle) like.
The Hamiltonian for the excitations close to the nodal points is
\begin{eqnarray}
H_D = \sum_{i,{\bf k},\sigma} \Psi^{\dag}_{i,\sigma}({\bf k}) 
\left[v_{F} k_{\perp,i} \sigma^z + v_{0} k_{||,i} \sigma^x \right] 
\Psi_{i,\sigma}({\bf k}) 
\label{dirac}
\end{eqnarray}
where $v_{F}$ is the Fermi velocity, $v_{0} = |\partial_k \Delta_{cdw}({\bf k})|$, 
$k_{\perp,i}$ ($k_{||,i}$) is the momentum perpendicular (parallel) to the Fermi surface . All these quantities are calculated at the Dirac points. Here $\sigma^a$ with $a=x,y,z$ are Pauli matrices which act in the spinor subspace. Hamiltonian (\ref{dirac}) describes a system of Dirac fermions with energy  $\epsilon^{\pm}_{i,{\bf k}}= \pm \epsilon_{i,{\bf k}}$ where
$\epsilon_{i,{\bf k}} = \sqrt{v^2_{F} k^2_{\perp,i} + v^2_{0} k^2_{||,i}}$. 
In what follows we drop this index $i$ since the Dirac electrons are decoupled.

The triple CDW transition breaks the inversion symmetry of the lattice leading to a three sublattice structure \cite{review}. In this case a piezo-electric coupling can develop between the Dirac fermions and acoustic phonons \cite{mahan}. The acoustic phonons are described by
\begin{eqnarray}
H_A = \frac{1}{2 \rho_L} \int d{\bf r} \left[ \Pi^2({\bf r}) + 
\rho_L^2 c_{s}^2 (\nabla \Phi({\bf r}))^2\right]
\label{phonons}
\end{eqnarray}
where $c_s$ is the sound velocity, $\rho_L$ is the lattice mass density and $\Pi({\bf r})$ and $\Phi({\bf r})$ are the momentum and phonon fields, respectively. The electron-phonon coupling is
\begin{eqnarray}
H_C =  \kappa   
\int d{\bf r} \, \Phi({\bf r}) \sum_{\sigma} \Psi^{\dag}_{\sigma}({\bf r})
\Psi_{\sigma}({\bf r})
\label{hc}
\end{eqnarray}
where $\kappa$ is the piezo-electric coupling constant. The electron Green's function for the problem described by Hamiltonian (\ref{dirac}), (\ref{phonons}) and (\ref{hc}) can be calculated self-consistently \cite{subir}. The main result of the interaction is the damping of these excitations. As a result $\Im[\Sigma(k_F,\omega)]$ is given in (\ref{marginal}) with $\tau_0^{-1}=27$ meV, $\alpha=2.14$ and $\gamma=0.212$. These results are valid for $\omega \ll Max[|\Delta_{cdw}|] \approx 60$ meV (obtained from the onset of the optical absorption \cite{leo} and ARPES \cite{kevin}). For $\omega > Max[|\Delta_{cdw}|]$ the Dirac electron description breaks down and $\Im[\Sigma(k_F,\omega)] \approx \tau_1^{-1} = 73$ meV. In Fig.\ref{photo} we plot our results against the experimental data. The agreement is remarkable given the simplicity of the model. Furthermore, the predictions of the marginal Fermi liquid phenomenology originally proposed for the HTC \cite{mfl} can now be extended to the case of MTD. In fact, recent dynamical transport measurements have pointed out the striking similarities between the MTD and HTC \cite{leo}.

Besides leading to damping, the phonons generate a retarded interaction. This retarded interaction, like in an ordinary superconductor, leads to pairing in the singlet channel. After tracing out the phonons the pairing Hamiltonian becomes:
\begin{eqnarray}
H_P = - g \sum_{{\bf k},{\bf k'}} \sigma^y_{a,b} \sigma^y_{c,d} \psi^{\dag}_{a,\uparrow}({\bf k'})
\psi^{\dag}_{b,\downarrow}(-{\bf k'}) \psi_{c,\uparrow}({\bf k}) \psi_{d,\downarrow}(-{\bf k})
\label{hp}
\end{eqnarray}
where $g \approx \kappa^2/\omega_D$ is the coupling constant ($\omega_D \approx c_s \Lambda$, with $\Lambda \approx 1/a$, is a Debye frequency). At the mean-field level the pairing Hamiltonian is written as
\begin{eqnarray}
H_{P} = \sum_{{\bf k},a,b} \left(\sigma^y_{a,b} \Delta_s  \psi^{\dag}_{a,\uparrow}({\bf k})
\psi^{\dag}_{b,\downarrow}(-{\bf k})
+ h.c. \right)
\label{hpmf}
\end{eqnarray}
where
\begin{eqnarray}
\Delta_s = - g \sum_{{\bf k},a,b} \sigma^y_{a,b} \langle \psi_{a,\uparrow}({\bf k})
\psi_{b,\downarrow}(-{\bf k}) \rangle 
\label{deltas}
\end{eqnarray} 
is the superconducting order parameter. The problem described by (\ref{dirac}) and (\ref{hpmf}) reduces to the diagonalization of a $4 \times 4$ matrix via a Bogoliubov transformation. The eigen-energies are: $E^{\pm}_{{\bf k}} = \pm \sqrt{\epsilon_{{\bf k}}^2 + |\Delta_s|^2}$. Thus, in the superconducting phase the Fermi surface is fully gapped by a CDW gap along the $\Gamma$-K direction and a superconducting gap along the $\Gamma$-M direction. $|\Delta_s|$ given by:
\begin{eqnarray}
|\Delta_s(T,g)| = 2 T \cosh^{-1}\left[\cosh\left(\frac{2 \pi v_F v_0}{T g_c}\right) e^{-\frac{2 \pi v_F v_0}{T g}}
\right]
\label{finaldelta}
\end{eqnarray}
where $g_c = 4 \pi^{3/2} \sqrt{v_F v_0}/\Lambda$. At $T=0$ we have:
\begin{eqnarray}
|\Delta_s(0,g)| = 4 \pi v_F v_0 \left(1/g_c-1/g\right)
\label{deltat0}
\end{eqnarray} 
which shows that superconductivity is only possible for $g>g_c$ at $T=0$. Thus, $g=g_c$ is a quantum critical point (QCP) and $g_c$ is the critical coupling constant. This result implies that there is a critical lattice spacing, $a_c$, below which superconductivity is not possible.
The critical mean-field temperature is $T^*(g) = |\Delta_s(0,g)|/(2 \ln(2))$. Notice that the dependence of the order parameter with the coupling constant is very different from the BCS expression (which does not require a critical coupling constant). This transition, as was noted earlier \cite{dima}, can be associated with the spontaneous chiral symmetry breaking in the Yukawa-Higgs model. In our case, however, the order parameter is complex implying that there is an extra U(1) symmetry associated with the phase of the order parameter. Because of the strong phase fluctuations in 2D the phase transition can only be of the KT type (2D-XY) due to the unbinding of vortex-anti-vortex pairs at a temperature
\begin{eqnarray}
T_{KT}(g) = \pi \sigma_s(g)/(2 m^*)
\label{tkt}
\end{eqnarray}
where $m^*$ is the pair effective mass and $\sigma_s(g) \propto |\Delta_s(0,g)|^2$ is the planar superfluid density. This transition, however, does not produce true long-range order. 
The weak coupling between layers changes the universality class of the transition to 3D-XY with $T_c \approx T_{KT} + b/\ln^2(T_{KT}/(c U_{\perp}))$ where $b$ is a number of order unit,
$U_{\perp}$ is the coupling energy per unit of length between layers, and $c$ is the inter-layer distance (we assume, $c \ll T_{KT}/U_{\perp}$). Thus, $T_c$ grows with decreasing $c$. As we can see the relevant parameter which controls the superconducting transition is $a/c$. In Fig.1 we plot the transition temperatures as a function of this parameter. We can clearly see the anti-correlation between $T_{cdw}$ and $T_c$. Notice that TaSe$_2$ with $T_c \approx 0.1$ K is close to the QCP. The growth of $T_c$ under pressure \cite{yamaya} is also consistent with our picture. Because $T^*>T_c$ we expect the a pseudo-gap region for $T_c < T < T^*$ where the order parameter is developed but there is no phase coherence. 

As we have shown there are many similarities between the problem discussed here and HTC.
Perhaps one of the most interesting is the anti-correlation between the $T_{cdw}$ and $T_c$. This is the same type of anti-correlation observed between the pseudo-gap energy scale and $T_c$ in the case of HTC \cite{loram}. The main difference, of course, is that at $T_{cdw}$ there is a true second order phase transition while the pseudo-gap temperature seems to be a crossover energy scale. In fact, the metallic state described here is similar in many respects to the nodal liquid description of HTC \cite{balents}. However, the MTD Dirac fermions are not related superconductivity but to a CDW state. In TMD the anomalous metallic behavior is due to the coupling to acoustic phonons (which are critical modes associated with the breaking of translation symmetry) while in the case of HTC this is not so clear. As shown in ref.\cite{subir}, phase fluctuations alone are not capable of producing marginal Fermi liquid behavior. In order to obtain this kind of behavior the system needs be close to a different type of QCP. There are many attempts to generate hidden QCP for HTC. Among them, the idea of stripes (a state closely related to CDW) have been discussed in the literature \cite{stripes}. It is worth noticing that TMD also show {\it stripe} phases associated with the breaking of the hexagonal symmetry in the triple CDW phase \cite{fleming}. Another possibility is related with the breaking of the time reversal symmetry and generation of gapped superconducting phases \cite{dima}. More recently it has been proposed that due to strong electron-electron correlations an exotic CDW state might be the hidden QCP of HTC \cite{sudip} and that the pseudo-gap temperature marks a real second order transition that is smeared by disorder. If this is indeed correct then the anomalies in these 2D systems can have the same origin. 

In summary, we have proposed a microscopic theory for the CDW and superconducting phases of TMD that involve the formation of a gapless CDW state. The elementary excitations are Dirac fermions that pair up due to the coupling to phonons and generate a superconducting state at low temperatures. We show that in the CDW state the coupling to phonons leads to marginal Fermi liquid behavior and to many anomalies in the physical properties. We show that our theory can explain the available experimental data including ARPES. We predict that high-resolution ARPES should be able to measure the modulation of the CDW gap around the Fermi surface and the opening of a superconducting gap at the Dirac points. The similarities between the phenomena described here and the one observed in other layered systems (such as HTC) is striking and the study of such exotic CDW might illuminate the way for the understanding of the complex physics in transition metal systems.

I acknowledge invaluable discussions with L.~Balents, 
D.~K.~Campbell, E.~Carlson, 
H.~Castillo, C.~Chamon, C.~McGuinness, C.~Nayak, S.~Sachdev and K.~Smith. 
I thank K.~Smith for providing the ARPES data.

$^*$On leave from Department of Physics, University of California, Riverside, CA 92521.

\begin{figure}
\epsfysize5 cm
\hspace{0cm}
\epsfbox{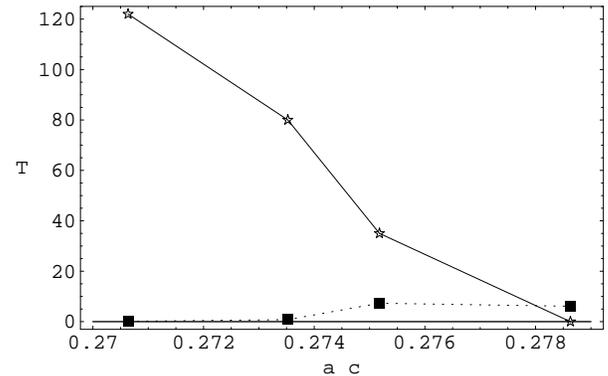}
\caption{Phase diagram: Stars, $T_{cdw}$; Filled squares, $T_c$. 
From left to right, TaSe$_2$, TaS$_2$, NbSe$_2$ and
NbS$_2$ \protect\onlinecite{exp1,exp2}. $a$ is the in-plane lattice
spacing and $c$ is the inter-layer spacing.}
\label{phasediagram}
\end{figure}

\begin{figure}
\epsfysize6 cm
\hspace{1cm}
\epsfbox{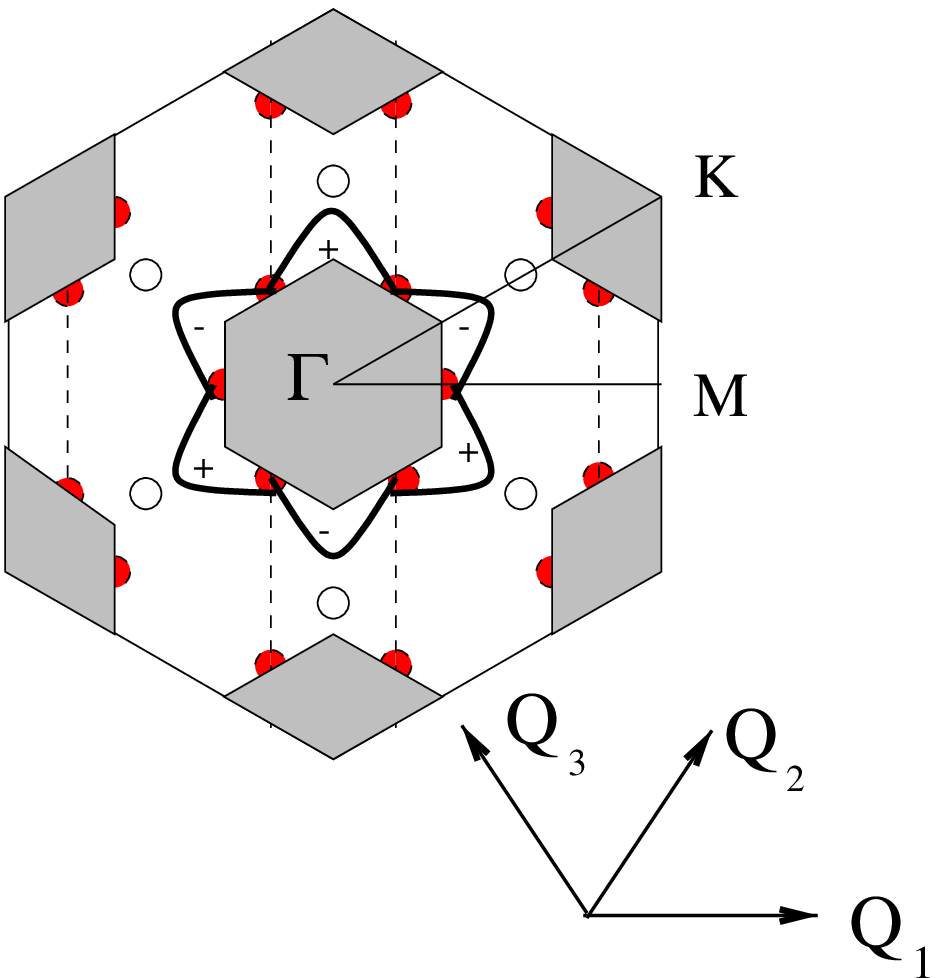}
\caption{Skematic plot of the Fermi surface accordingly to 
\protect\onlinecite{wexler}. Dashed lines: nodal lines associated
with ${\bf Q}_1$;
Filled circles: Dirac points; Empty circles: Saddle points;
Thick line: proposed CDW gap.}
\label{fermisurface}
\end{figure}

\begin{figure}
\epsfysize5 cm
\hspace{0cm}
\epsfbox{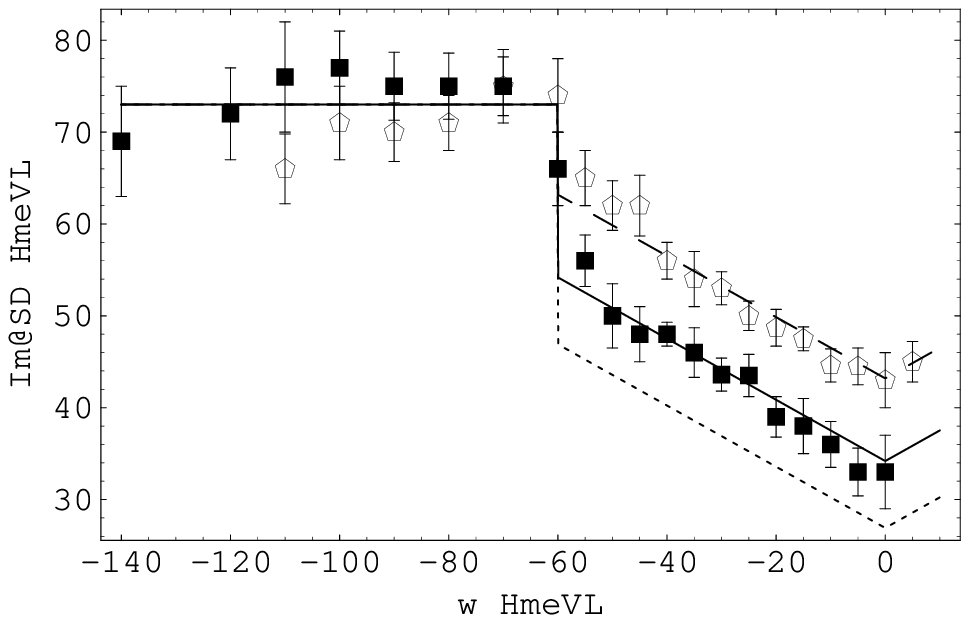}
\caption{$\Im[\Sigma(k_F,\omega)]$. Experiments: squares ($T=34$K), pentagons ($T=76$K) \protect\onlinecite{kevin}; Theory: dotted ($T=0$), continous ($T=34$K), dashed ($T=76$K).}
\label{photo}
\end{figure}

\end{document}